\documentclass[]{spie}  
\usepackage[]{graphicx}
\usepackage{siunitx}
\usepackage{caption}	
\usepackage{subcaption}	
\usepackage{sidecap}
\usepackage{wrapfig}

\DeclareSIUnit[per-mode=symbol]{\cms}{\cm\per\second}
\DeclareSIUnit[per-mode=symbol]{\mps}{\meter\per\second}
\DeclareSIUnit{\micron}{\micro\meter}
\title{Characterizing octagonal and rectangular fibers for MAROON-X} 

\author{Adam P. Sutherland\supit{a,b}, Julian Stuermer\supit{a}, Katrina R. Miller\supit{a,c}, Andreas Seifahrt\supit{a}, \\and Jacob L. Bean\supit{a}
\skiplinehalf
\supit{a}University of Chicago, USA;
\supit{b}University of Arizona, USA;
\supit{c}Duke University, USA
}

\authorinfo{Further author information: (Send correspondence to Andreas Seifahrt: E-mail: seifahrt@uchicago.edu, Telephone: 1 773 702 9877)}

\begin{document} 
  \maketitle 

\begin{abstract}
We report on the scrambling performance and focal-ratio-degradation (FRD) of various octagonal and rectangular fibers considered for MAROON-X. Our measurements demonstrate the detrimental effect of thin claddings on the FRD of octagonal and rectangular fibers and that stress induced at the connectors can further increase the FRD. We find that fibers with a thick, round cladding show low FRD.
We further demonstrate that the scrambling behavior of non-circular fibers is often complex and introduce a new metric to fully capture non-linear scrambling performance, leading to much lower scrambling gain values than are typically reported in the literature ($\leq$1000 compared to
10,000 or more). We find that scrambling gain measurements for small-core, non-circular fibers are often speckle dominated if the fiber is not agitated.
\end{abstract}



\section{INTRODUCTION}
\label{sec:intro}  

MAROON-X is a new fiber-fed, red-optical, high-precision radial-velocity spectrograph for one of the twin 6.5m Magellan Telescopes in Chile. The instrument is currently under construction at the University of Chicago\cite{seifahrt1}. MAROON-X will be fed by a \SI{100}{\micron} octagonal fiber at $f$/3.3 from the telescope, delivering a FOV of 0.95'' on sky. To achieve a resolving power of ~80,000, a micro-lens based pupil slicer and double scrambler\cite{seifahrt2} reformats the light into three \SI{50 x 150}{\micron} rectangular fibers at $f$/5.0, which are then close stacked to form a pseudo-slit at the spectrograph input. 

We have tested the scrambling performance and focal-ratio-degradation (FRD) of a number of octagonal and rectangular fibers to select the best fibers for this application.

\section{MEASUREMENT SETUP}
\label{sec:setup}
Since we needed to test fibers with comparatively small core dimensions, we required a setup capable of imaging extremely small pinholes onto fiber faces with good optical quality. We have thus designed a stable test setup based on 2'' off-axis-paraboloids (OAP) to facilitate FRD and scrambling gain measurements with pinholes as small as \SI{5}{\micron} with selectable $f$-ratios as large as $f$/2.0 at different wavelengths and band passes. The setup also allows to insert pupil masks to simulate obstructions, like telescope secondaries, and to image the input pupil. 

\begin{figure}[t!]
\centering
\includegraphics[trim={0cm 3cm 0cm 3cm},width=1.0\linewidth,clip]{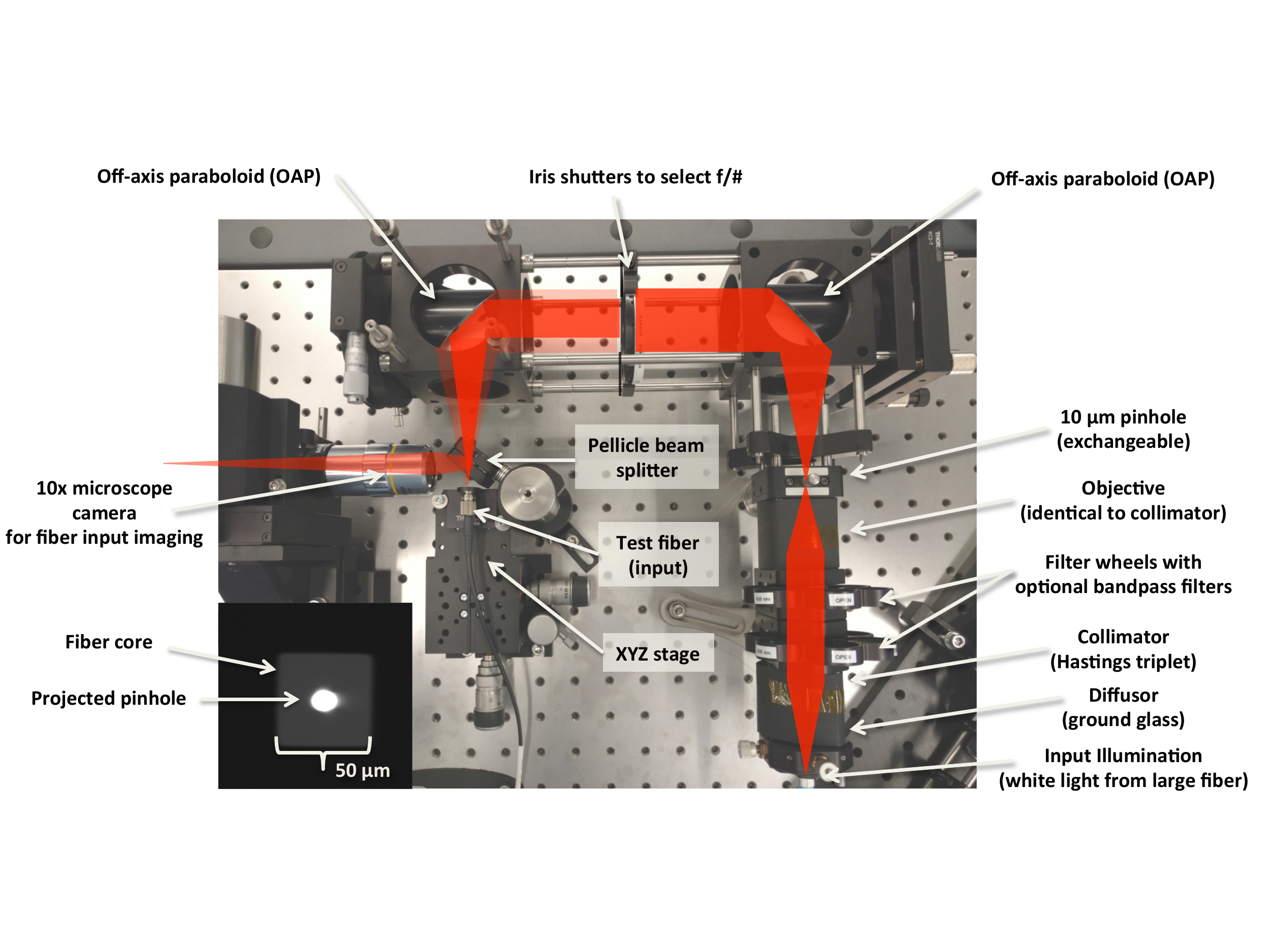}
\vspace{1mm}
\caption{\textbf{Input arm of our fiber test station.} The optical path is indicated as the red overlay. The setup illuminates a pinhole with light from a laser-driven white light source (LDLS) and re-images the pinhole with a selectable $f$-ratio onto the fiber input face. Image (pinhole) size and shape, wavelength and bandpass, as well as the pupil illumination ($f$-ratio via an iris, obstructions via masks) can be adjusted on the fly. The pinhole image is projected on the test fiber which can be moved precisely on a XYZ flexure stage and simultaneously imaged with a 10x microscope objective. An example of a \SI{10}{\micron} pinhole projected on a \SI{50x50}{\micron} rectangular fiber is shown in the insert at the lower left. The horizontal elongation of the pinhole   image is caused by the pellicle beam splitter and is not present on the fiber input face. The left OAP can be rotated by 90 degree to steer the beam upwards. By placing the far-field camera on top the OAP cage, the input pupil can be imaged.}
\label{fig1}
\end{figure}

A pellicle beam splitter in front of the fiber redirects 5\% of the light to a 10x microscope objective and CMOS camera setup to image the input face of the fiber during the test procedure. A Thorlabs flexure stage with differential micrometers allows to move the fiber position relative to the pinhole image projected on the fiber face with low crosstalk and sub-\si{\micron} precision and repeatability (see Fig.\,\ref{fig1}). The XYZ stage supports both connectorized fibers and bare fibers held in a v-groove.

\begin{wrapfigure}[37]{r}{0.45\textwidth}
      \centering
        \includegraphics[trim={6.3cm 0.5cm 7.0cm 0cm}, clip,width=1\linewidth]{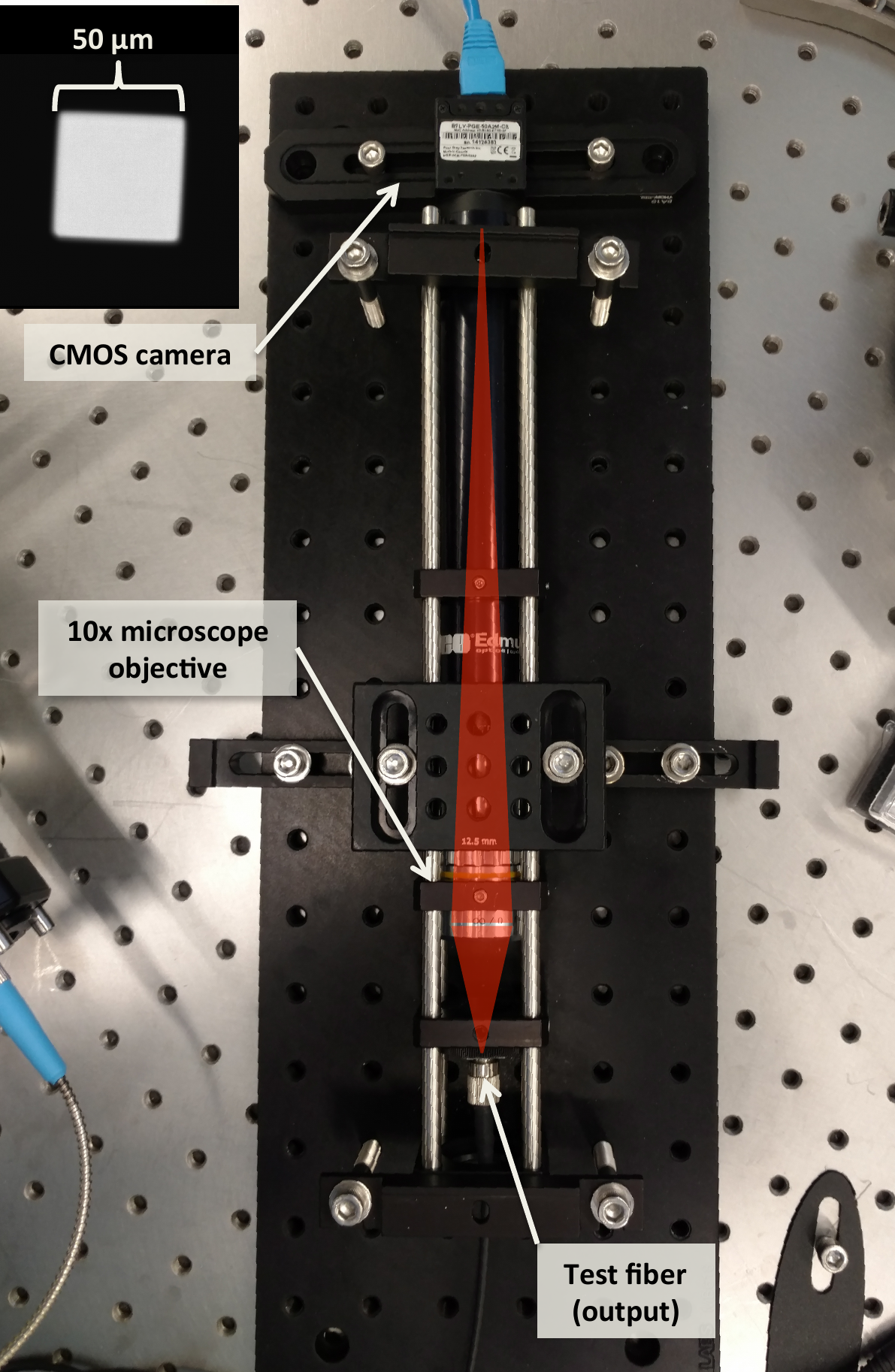}
\vspace{1mm}
      \caption{\textbf{Near-field camera setup.} Near-field images of fibers illuminated by the input setup (see Fig.\,\ref{fig1}) are recorded with a pixel scale of \SI{0.22}{\micron\per pixel} on a CMOS camera. No moving parts and an additional breadboard improve the thermo-mechanical stability of the setup. An example of a near-field image of a \SI{50x50}{\micron} rectangular fiber is shown in the insert in the upper left corner. The input illumination is shown in Fig.\,\ref{fig1}.}
      \label{fig2}
\end{wrapfigure}

Near-field images are recorded with a second CMOS camera and 10x microscope objective, mounted on a separate breadboard for increased thermo-mechanical stability (see Fig.\,\ref{fig2}). Test fibers need to be connectorized or inserted in a bare fiber terminator to be imaged.

Far field images are obtained with a Thorlabs Hastings triplet and a SBIG STF-8300 CCD camera. Zemax image simulations allow a direct conversion from measured pupil size in the far-field image to $f$-ratio. 

A second setup of two OAP collimators and an iris shutter, similar to the input setup shown in Fig.\,\ref{fig1}, allows photometric FRD measurements. The setup is tightly shielded against stray light to keep photometric errors at or below 1\%. We use a custom tool to select identical apertures at the input and output iris (i.e., identical $f$-ratios) with accuracies better than 3\%. A Thorlabs PM100 power meter is used to measure the flux at the position of the re-imaged fiber face.

\section{METHODS AND METRICS}
\label{sec:methods} 
The scrambling performance of a fiber is typically characterized by the scrambling gain (SG). The scrambling gain is traditionally defined as the ratio between the relative displacement of the star, simulated as a pinhole, in front of the input fiber face and the corresponding shift of the point spread function (PSF) on the focal plane of the spectrograph\cite{Avila06,Avila08}:
\begin{equation}
SG = \frac{(d_{input}/D_{input})}{(f/F)} = \frac{(d_{input}/D_{input})}{(d_{output}/D_{output})}
\label{eq:sg}
\end{equation}
where $D_{input}$ is the fiber diameter, $d_{input}$ the shift of the star (or pinhole) , $f$ the shift of the PSF and $F$ the FWHM of the PSF. $f/F$ can be replaced by the shift in the barycenter ($d_{output}$) and the width of the near-field fiber face image ($D_{output}$), respectively. The latter allows the characterization of the fiber independent of the spectrograph.
Since the definition of the scrambling gain relies on a linear relation between input and output image shift, the strongest signals are expected when the pinhole image is placed at or near the edge of the fiber core, effectively reducing the amount of measured points to two\cite{Halverson}. 
However, non-circular fibers with high scrambling gains often exhibit a non-linear behavior between input image displacement and the barycenter shift of the near-field\cite{Chazelas10,Feger12}. We speculate that this is due to residual modal (speckle) effects even for wide optical bandpasses as well as asymmetries caused by impurities in the fiber and stress at the connectors.

We thus modified Equ.\,\ref{eq:sg} to remove the artificial linear coupling between in- and output shifts and define a new metric for measuring scrambling called $SG_{min}$. $SG_{min}$ uses the maximum variation in near-field barycenter movement divided by the maximum change of input position as defined by equation \ref{eq:sgmin}:
\begin{equation} 
SG_{min} = \frac{\max(d_{input}/D_{input})}{\max(d_{output}/D_{output})} = \frac{D_{input}}{\max(d_{output})}
\label{eq:sgmin}
\end{equation} 
with the same annotations as equation \ref{eq:sg} and under the assumption of sampling the complete width of the fiber input face.

This metric uses the maximum barycentric movement of the near-field image due to any change in the input pinhole position to determine the scrambling of the fiber. It thus returns a lower limit of the scrambling gain, hence the notation as $SG_{min}$. Since this metric takes all input locations across the fiber face into consideration it does not return spurious high scrambling gain values based on measurements of only a few input illumination positions. $SG_{min}$ will return the same result as the classical scrambling gain in case of a strictly linear relationship of input illumination and near-field image shift, but delivers a more realistic metric for more complex relationships. In order to make $SG_{min}$ robust against outliers, we discard 2.5\%  of the most extreme near-field barycenter positions at either end of the distribution. 

Due to the ambiguity in orientation between the input and output face for octagonal and square fibers, we measure the barycentric shift in $X$ and $Y$ and combine them into a single metric per fiber:  
\begin{equation}
<SG_{min}> = (SG_{min,X}^{-2}+SG_{min,Y}^{-2})^{-\frac{1}{2}}.
\end{equation}

Scrambling measurements were taken by repeatedly moving the pinhole image across the input face, slowly scanning the complete face of the fiber in one direction multiple times. 200--300 images of the fiber input and corresponding near-field images are recording in two video streams. The images are later re-indexed using an image time stamp. Scanning over the input fiber face multiple times in each test helps to average over mechanical vibrations and reveals potential correlations with drifts of the physical location of the fiber and the camera positions. The exposure time of each frame is dependent on the wavelength and $f$-ratio of the light as well as the size of the fiber. It ranges from tens of milliseconds to greater than one second. The duration of the measurement is dependent on the exposure time of the individual frames and lasts between one and five minutes. 

Vibrations and mechanical drifts in the setup can move the image of the output face of the fiber on the detector, acting like apparent barycentric changes in the near-field. In order to separate actual changes in the near-field illumination of the fiber from the physical movement of the fiber, we trace the physical location of the fiber throughout the duration of the measurement. After dark subtraction, we apply custom software tools in Python to determine the position of the fiber face by constructing a mask from thresholding a median filtered version of each near-field image in order to find the outline of the fiber face. We measure its position independent of the light distribution within the fiber core and subtract the change of the mask position from the change in the barycenter of the image to remove physical image drifts. 

Correcting the near-field images for spurious movements of the fiber removes both high frequency vibrations and lower frequency drifts, but is limited by our inability to track vibrations smaller than a few nanometers or frequencies shorter than the exposure time of the image. $SG_{min}$ measurements are thus ultimately noise-limited for fibers with very high scrambling where the spread of the near-field shifts due to noise and instabilities approaches the maximum barycentric shift due to changes in the input illumination. Careful measurements for static pinhole positions with white light illumination and small $f$-number, both suppressing the speckle pattern, indicate a upper limit for $SG_{min}$ values of 35,000--50,000. The majority of fibers tested had near-field barycentric movements much greater than this noise limit. 

\subsection{Focal-Ratio-Degradation}

Focal-ratio-degradation (FRD) has been studied intensively in the literature and is understood to depend on a variety of factors, such as sensitivity of the fibres to mechanical stresses, particularly at fibre ends (connectors), scattering processes, as well as optical alignment errors\cite{Avila06,Haynes11}.
Stresses induced by the shrinkage of adhesive in the ferrules when terminating the fibers seem to play a particularly important role\cite{Avila12}.

In order to minimize FRD effects for MAROON-X, we opted to operate the main octagonal fiber link at $f$/3.3 and keep the short rectangular fiber link between the pupil slicer and the spectrograph slit at a moderate $f$/5.0, leaving the final conversion to the $f$/10 acceptance cone of the spectrograph to a re-imaging relay. We have thus tested the FRD of our fibers for input illuminations of $f$/2.5, $f$/3.5 and $f$/5.0 only. 

We used two methods to look at the far-field behavior. First, we project a pinhole at various input $f$-ratios onto the input face of the fiber (as we did for our scrambling tests) and imaged the far-field with and without pupil masks that create a central obstruction. We find that the far-field scrambling of octagonal and rectangular fibers shows the same behavior as for circular fibers - good azimuthal scrambling but poor (or a complete lack of) radial scrambling. The image of a central circular pupil obstruction is thus preserved in the absence of FRD and cladding modes\cite{Avila10}. FRD is caused by the scattering of light, which works in both directions, out of the input cone and into the input cone. The latter effectively fills a dark central pupil obscuration (high $f$-number angle space) with light. This effect is typically not captured by FRD measurements as the scattered rays stay within the input cone and are thus not lost under a fixed output $f$-ratio constraint. However, the softening of the sharp transition at the edge of the central obscuration is easier to observe than the added light at the outer edge of the pupil, which is often intrinsically softer and less well defined. Small scattering from added stress, either by bending the fiber or slightly pressing on the bare fiber in a v-groove or the fiber terminator can thus be easier detected using the amount of light in the central obscuration as a qualitative, yet very sensitive, indicator of FRD. 

While quantitative FRD measurements can be extracted from far-field images without input pupil obscuration, we opted instead to use photometric measurements where the input and output $f$-ratios are selected on an iris shutter with a sufficient degree of repeatability. We have compared a number of FRD values obtained from the photometric setup and from analysis of the far-field images and find a satisfying level of agreement. Since our desired input $f$-ratios are fixed, we show FRD plots as the encircled energy for a given output $f$-number, normalized to $f$/2 -  the maximum $f$-number our setup can support. Our FRD numbers thus ignore absorption and Fresnel losses. 

\section{OCTAGONAL FIBERS}\label{sec:octagonal} 

We have characterized two octagonal fibers with \SI{100}{\micron} core diameter, a EFBP fiber from Polymicro and a WF fiber from CeramOptec. The Polymicro fiber was purchased as a factory prepared \SI{7}{\meter} patch cord with FC/PC connectors. The CeramOptec fiber was purchased as bare fiber and a \SI{10}{\meter} patch cord was produced by us using FC/PC connectors and ultra-low shrinkage adhesive. 

\subsection{Focal-Ratio-Degradation}\label{sec:octagonal-frd}

\begin{figure}[t!]
\centering
\includegraphics[trim={0cm 0.3cm 0cm 0.3cm},width=1.0\linewidth,clip]{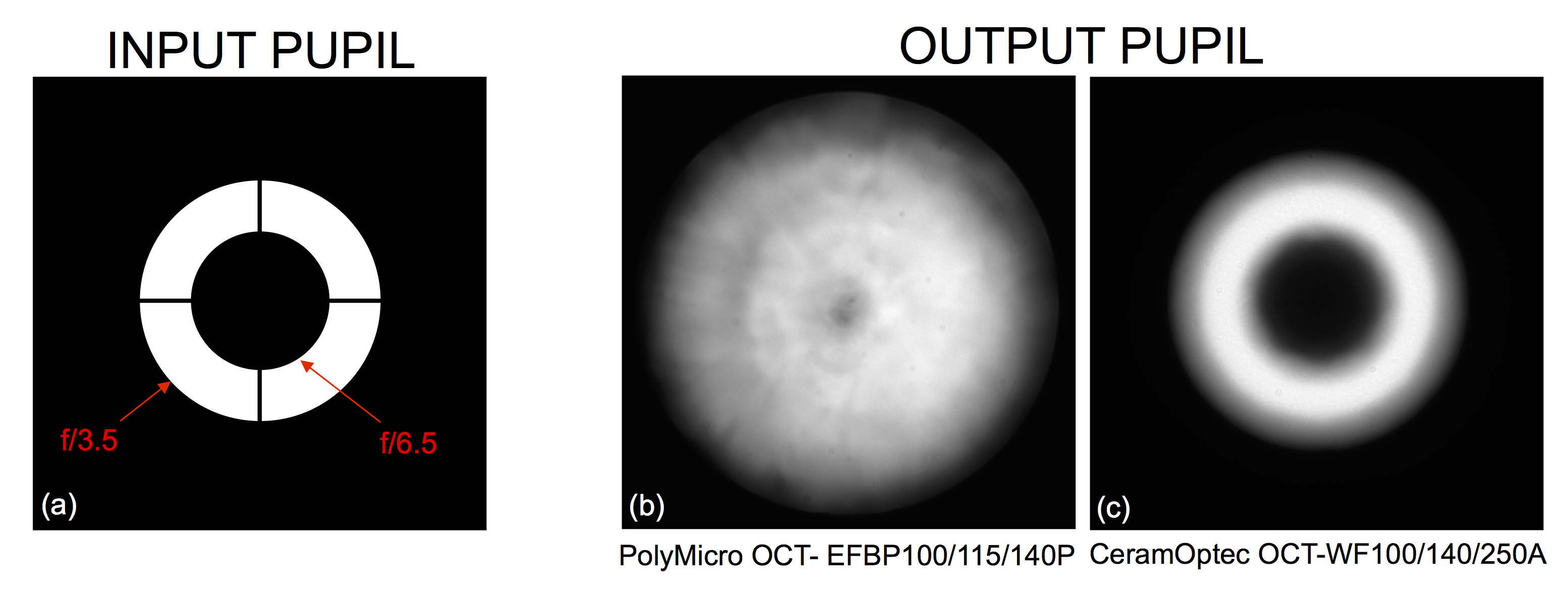}
\vspace{1mm}
\caption{\textbf{Example of far field images obtained for two octagonal fibers} with a $f$/3.5 input illumination and a $f$/6.5 obstruction. (a) Input pupil. (b) Far-field image at the output of a \SI{7}{\meter} long octagonal fiber from Polymicro. (c) Far-field image for a \SI{10}{\meter} long octagonal fiber from CeramOptec. FRD measurements for unobstructed pupils are shown in Fig.\,\ref{fig:oct-frd} below.}
\label{fig:oct-pupil}
\end{figure}

Inspection of far-field images readily revealed a strong difference in FRD (see Fig.\,\ref{fig:oct-pupil}). The Polymicro fiber showed extreme FRD, with little to no change in the far field when changing the input $f$-number or placing a pupil mask in the input beam. In contrast, the CeramOptec fiber showed a well behaved far field image with low apparent FRD. Both fibers where coiled up in a \SI{0.2}{\meter} diameter loop during the measurements.

\begin{figure}[b!]
\centering
\includegraphics[trim={1cm 0cm 2cm 0cm},width=0.49\linewidth,clip]{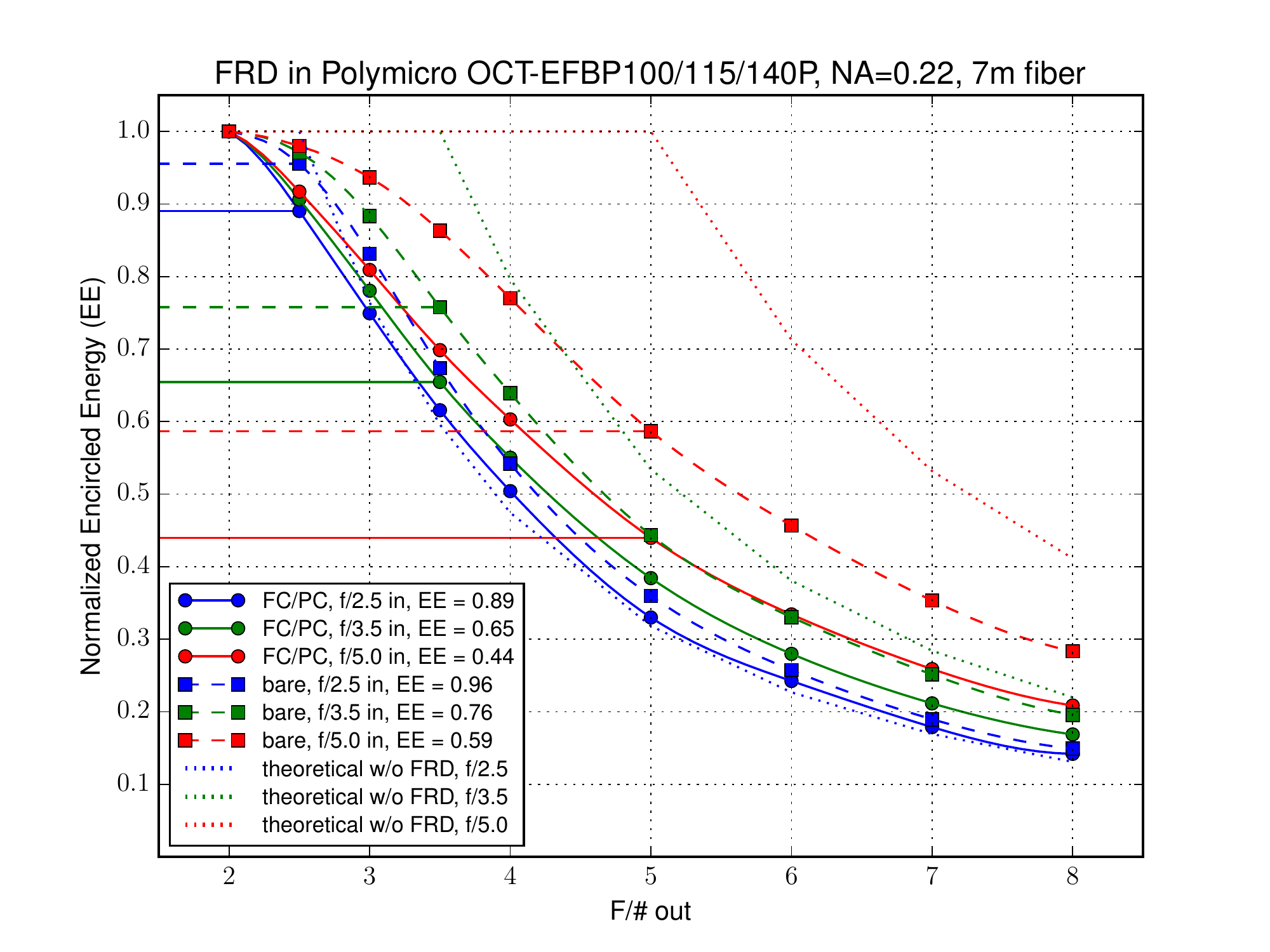}
\includegraphics[trim={1cm 0cm 2cm 0cm},width=0.49\linewidth,clip]{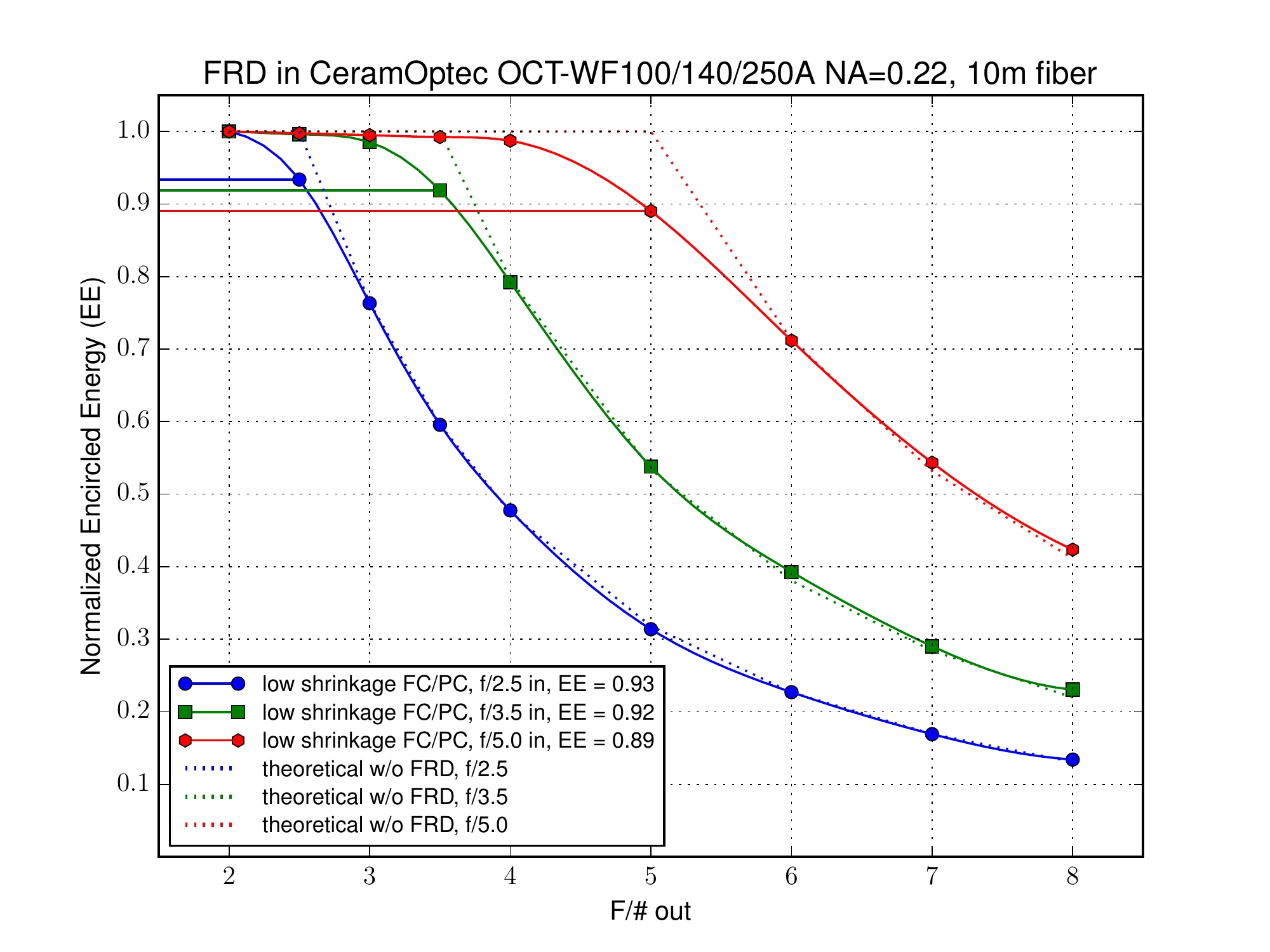}
\vspace{1mm}
\caption{\textbf{FRD measurements for two octagonal fibers from Polymicro (left) and CeramOptec (right)}. The \SI{100}{\micron} Polymicro fiber shows extreme FRD and increased light losses, both for a \SI{7}{\meter} connectorized fiber and a bare fiber of almost the same length. We hypothesize that the extremely thin cladding is the culprit, although we observe similar behavior in a larger \SI{200}{\micron} version of this fiber.}
\label{fig:oct-frd}
\end{figure}

To further investigate the reason for the strong FRD in the Polymicro fiber, we cut the connectors off and re-polished the bare fiber ends using Thorlabs bare fiber terminators. Re-examination of the now bare fiber revealed only a slight improvement in the FRD. We noticed a very high sensitivity of the bare fiber to any mechanical stress. When placing the bare fiber end in a v-groove, the FRD notably improved for moments when no mechanical constraint was used to keep the fiber in place. Even light pressure, either from the clamp of the bare fiber terminator or a magnetic fixture on the v-groove dramatically increased the FRD.

Quantitative FRD measurements show the same picture (see Fig.\,\ref{fig:oct-frd}). A low level of FRD is present in the CeramOptec fiber (89\% EE for $f$/5.0 in = out). In constrast, the Polymicro fiber showed strong FRD for the connectorized version (44\% EE) with only slight improvement for the bare fiber (59\% EE). We see the main reason for this behavior in the extremely thin cladding (\SI{115}{\micron}) and buffer (\SI{140}{\micron}) of the Polymicro fiber compared to the much thicker cladding (\SI{140}{\micron}) and buffer (\SI{250}{\micron}) of the CeramOptec fiber. 

\subsection{Scrambling}\label{sec:oct-scrambling}

\begin{figure}[h!]
\centering
\includegraphics[trim={0cm 0cm 0cm 0cm},width=1.0\linewidth,clip]{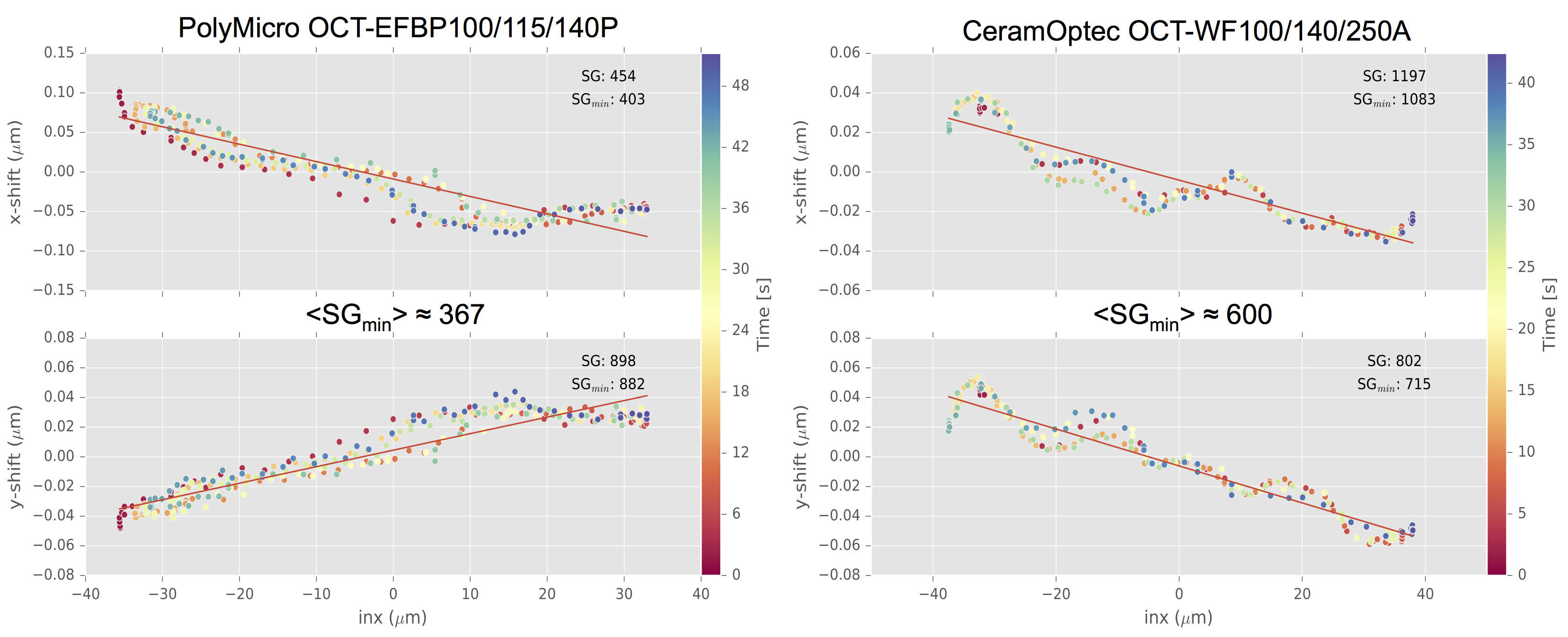}
\vspace{1mm}
\caption{\textbf{Scrambling gain measurements for two octagonal fibers with white light illumination} at $f$/3.5. The classical, linear scrambling gain is shown as a linear fit and matches our $SG_{min}$ metric. The image barycenter of the output fiber face was measured in X (top) and Y (bottom) while sweeping the location of a \SI{10}{\micron} pinhole multiple times across the input face of the fiber. Due to the unknown relative orientation of fiber input and output, the mean scrambling gain $<SG_{min}>$ is the inverse average of the measurements in X and Y. 
}
\label{fig:oct-scrambling}
\end{figure}

Scrambling gain measurements have been obtained for both octagonal fibers with small (\SI{10}{\micron}) pinhole images in white light illumination at $f$/3.5, close to the desired $f$/-number for the MAROON-X fiber feed. We find a reasonably linear behavior of the barycenter shift in the near-field vs. the pinhole location. Moderate scrambling gains of close to 400 for the Polymicro fiber and close to 600 for the CeramOptec fiber were obtained (see Fig.\,\ref{fig:oct-scrambling}). Even in white light, we see imprints of the modal structure in the CeramOptec fiber, evident as the regular pattern during the pinhole scan. This pattern is not dominating our scrambling gain measurements, but illustrates the presence of residual speckles in the near-field images even when averaged over wide bandpasses, an effect previously noticed only for narrower bandpasses of \SIrange{50}{100}{\nano\meter}\cite{Feger12}.

\section{RECTANGULAR FIBERS}\label{sec:rect} 
We have tested multiple rectangular fibers from CeramOptec, LEONI/j-fiber, and Mitsubishi with aspect ratios ranging from 1:1 to 1:4 and widths ranging from \SIrange{25}{57}{\micron}.
Except for the CeramOptec WF 50$\times$150/300 fiber, all fibers were tested as factory provided patch cords with lengths ranging from \SIrange{2}{5}{\meter}. We present here a representative sub-sample of the analyzed fibers.

\subsection{Focal-Ratio-Degradation}\label{sec:rect-frd}

Inspection of far-field images obtained for pupil illuminations as shown in Fig.\,\ref{fig:oct-pupil}, showed a wide range of FRD performance, with the three CeramOptec fibers showing the worst FRD behavior. Generally we found fibers with thin, rectangular claddings to show strong FRD effects.  

As for the octagonal fiber from Polymicro, we cut off the connectors from one of the patch cords (CeramOptec 33$\times$100/53$\times$120) and re-polished the bare fiber.
\begin{figure}[t!]
\centering
\includegraphics[trim={0cm 0.2cm 0cm 0.3cm},width=1.0\linewidth,clip]{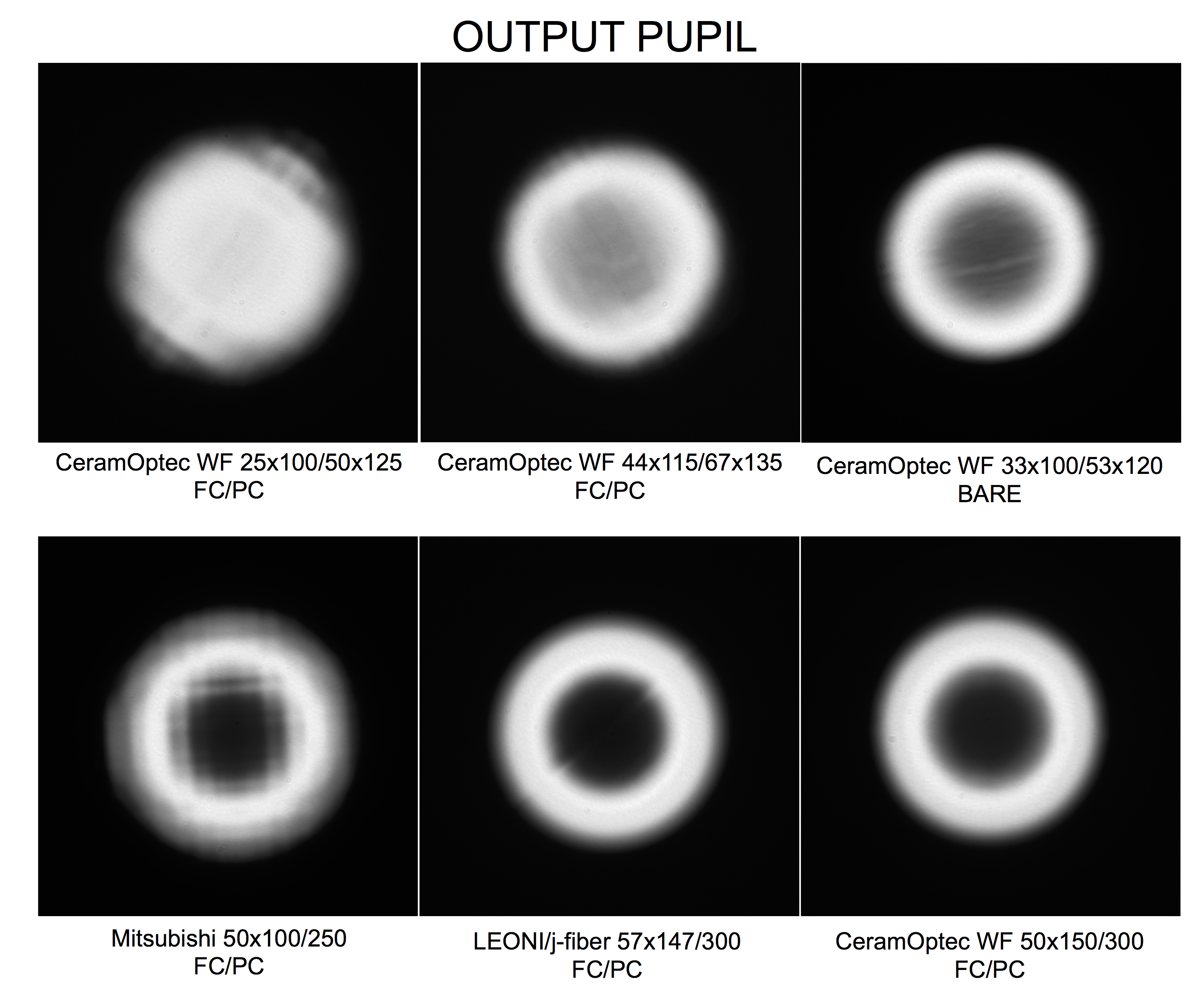}
\vspace{1mm}
\caption{\textbf{Example of far field images obtained for various rectangular fibers} with a $f$/3.5 input illumination and $f$/6.5 obstruction. See Fig.\,\ref{fig:oct-pupil} for an image of the input pupil. Fibers with thin, rectangular cladding (top row) show increased FRD and cladding modes. Fibers with thicker, round cladding (bottom row) show improved FRD behavior. FRD measurements for unobstructed pupils are shown in Fig.\,\ref{fig:rect-frd} below.
}
\label{fig:rect-pupil}
\end{figure}
We found a strong improvement in the FRD behavior of the bare fiber compared to the jacketed version. The sensitivity to mechanical stress was less pronounced compared to the octagonal fiber, likely due to the thicker buffer (\SI{250}{\micron}) of the rectangular fiber. However, we still see an increase in FRD when the fiber is fixed on a v-groove or held clamped in a bare fiber terminator. Only when completely unconstrained do we see a clean un-illuminated central obscuration in the far-field.

It is noteworthy that rectangular fibers with very sharp corners show regular patterns in the far-field (see, e.g., the Mitsubishi fiber in the lower left corner of Fig.\,\ref{fig:rect-pupil}).

\begin{SCfigure}[][t!]
\centering
\vspace{0mm}
\includegraphics[trim={1cm 0cm 2cm 0.5cm},width=0.65\linewidth,clip]{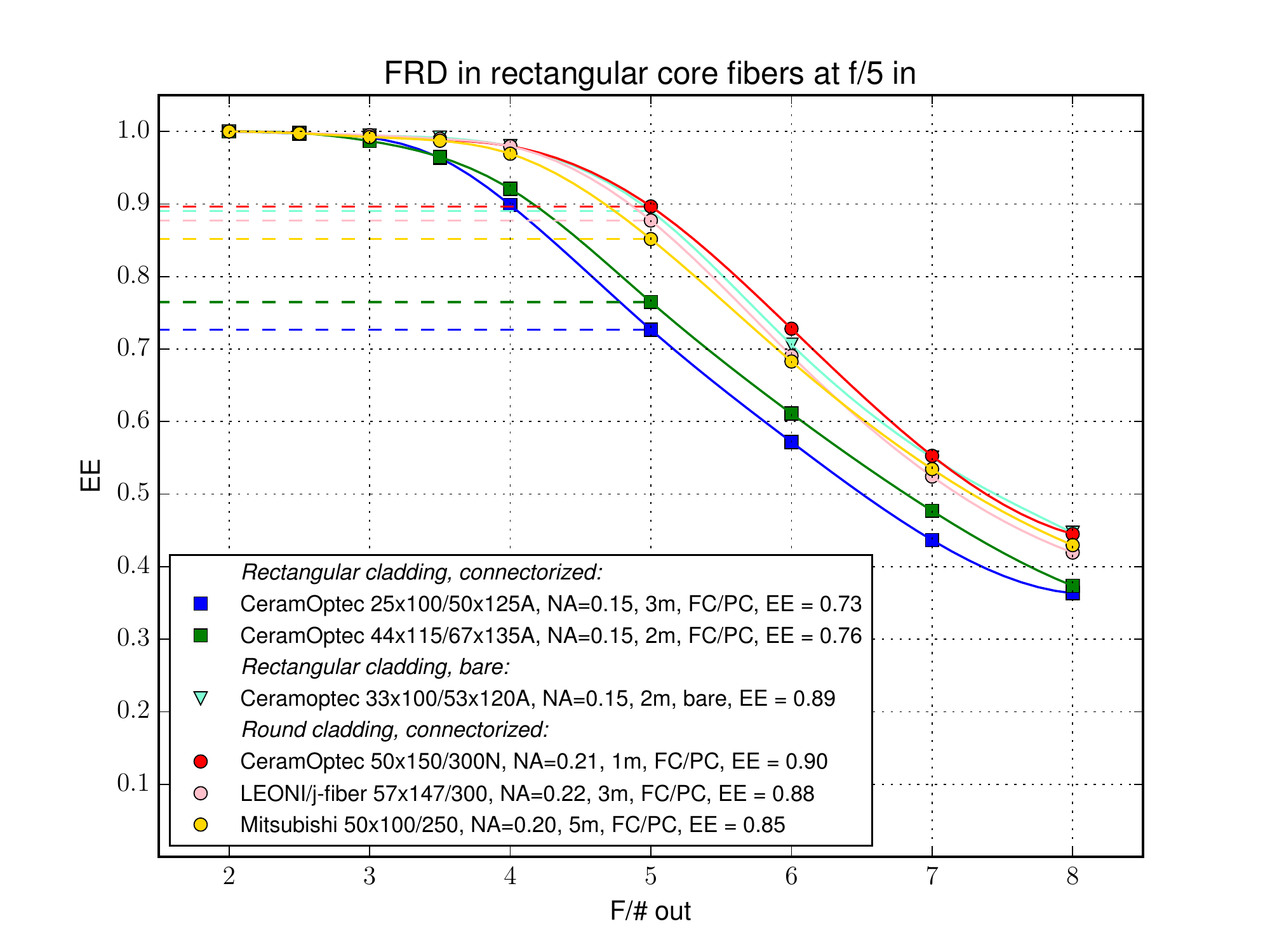}
\vspace{0mm}
\caption{\textbf{FRD measurements for various rectangular fibers} at $f$/5.0. The difference between fibers with thin, rectangular cladding and thick, round cladding is readily apparent. Removing the connectors from fibers with rectangular cladding greatly reduces stress induced FRD (see bare CeramOptec 33$\times$100 fiber vs. connectorized CeramOptec 44$\times$115).
}
\label{fig:rect-frd}
\end{SCfigure}

Quantitative FRD measurements were obtained with our photometric setup and we show the FRD values for a $f$/5.0 input illumination in Fig.\,\ref{fig:rect-frd}. As expected from the pupil images shown Fig.\,\ref{fig:rect-pupil}, the FRD for connectorized fibers with thin rectangular cladding is notably worse (EE=73\%--76\%) compared the bare version (EE=89\%) or to fibers with thicker, round cladding (EE=85\%--90\%). The fiber matching our geometrical constraints the best (CeramOptec 50$\times$150/300) shows one of the best FRD values for the desired $f$/5.0 illumination (EE=90\%).

\subsection{Scrambling}\label{sec:rect-scrambling}

\begin{figure}[t!]
\centering
\includegraphics[trim={0cm 0cm 0cm 0cm},width=1.0\linewidth,clip]{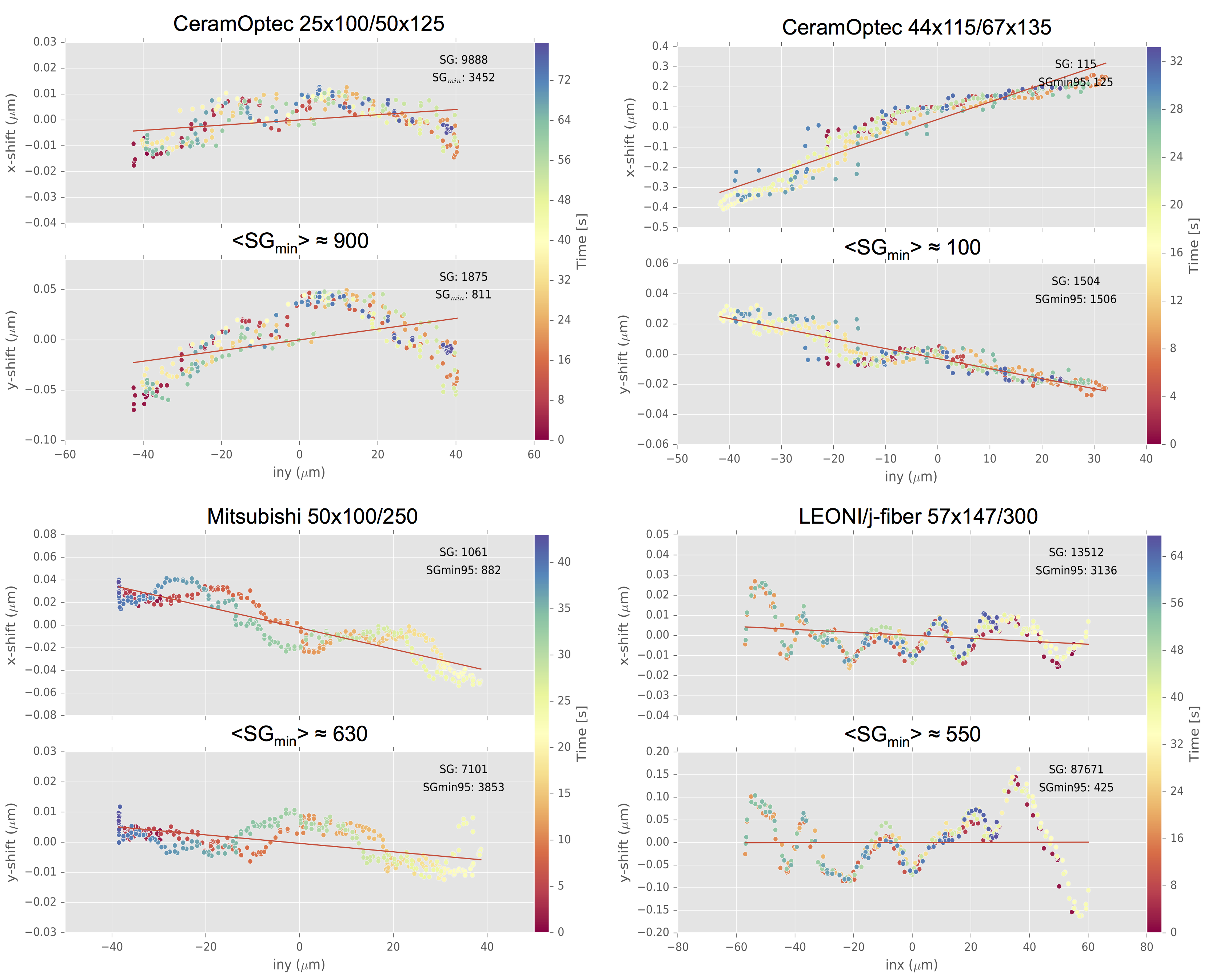}
\vspace{1mm}
\caption{\textbf{Examples of scrambling gain measurements for four rectangular fibers} at $f$/4.0 and $\lambda$=\SI[separate-uncertainty=true]{600+-20}{\nano\meter}. The classical, linear scrambling gain is shown as a linear fit and matches our $SG_{min}$ metric in cases of linear behavior of input vs. output barycenter shift. Note the overestimation of the scrambling gain for the highly non-linear behavior of the LEONI fiber (lower right panel). Data were obtained with the same procedure as for the octagonal fibers in Fig.\,\ref{fig:oct-scrambling}. $<SG_{min}>$ values are from multiple measurements at $f$/4 and $f$/6 between \SI[separate-uncertainty=true]{450+-20}{\nano\meter} and \SI[separate-uncertainty=true]{600+-20}{\nano\meter}.
}
\label{fig:rect-scrambling}
\end{figure}

Scrambling gain measurements were taken in the same way as for the octagonal fibers (see Sect.\,\ref{sec:oct-scrambling}). We show four representative examples in Fig.\,\ref{fig:rect-scrambling}. In all cases, the pinhole image was scanned over the long axis of the rectangular core. Some fibers showed pronounced deviation from a linear relationship between pinhole location and barycenter position. The CeramOptec 25$\times$100/50$\times$125 shows symmetric behavior around the center position which would lead to very high scrambling gain measurements if only data from pinhole locations near the edges of the core were recorded (upper left panel in Fig.\,\ref{fig:rect-scrambling}). Likewise, the LEONI/j-fiber 57$\times$147/300 fiber showed very pronounced speckle effects which dominate our $SG_{min}$ measurement but are not captured with a linear fit, leading to a 30$\times$ higher classical scrambling gain measurement compared to our $SG_{min}$ metric. Average $<SG_{min}>$ values for all rectangular fibers measured so far range from 100--1000, with a clear dependence on wavelength and $f$-number, further emphasizing the influence of modal structure on scrambling gain measurements. We have yet to measure the scrambling gain
of the CeramOptec 50$\times$150/300 fiber selected to feed the light from the pupil slicer of MAROON-X to the slit of the spectrograph\cite{seifahrt2}.

\section{SUMMARY}
\label{sec:discussion}
We have obtained near-field and far-field images as well as photometric FRD measurements of available octagonal and rectangular fibers of different sizes and from different manufacturers to select the best fibers for MAROON-X.

We find that far-field images of fiber with input pupils containing a central obscuration quickly reveal FRD and cladding modes due to the sharp transition between the central dark area and the bright ring in a FRD-free far-field image.

For both octagonal and rectangular fibers we find that thin claddings increase losses and make fibers more vulnerable to mechanical stresses. The latter appears also true for fibers with thin buffers. 

Rectangular claddings produce high FRD due to asymmetric stresses in standard connector ferrules, even with low-shrinkage (0.4\%) adhesive. Removing this stress (i.e., working with bare fibers) greatly improves the FRD behavior but poses practical limits on the usage of these fibers.

Scrambling behavior is often complex and our $SG_{min}$ metric captures the full non-linear scrambling behavior, leading to much lower scrambling gain values than are typically reported in the literature. Scrambling gain measurements for small-core, non-circular fibers are often speckle dominated if the fiber is not agitated.

\acknowledgments     
 
The University of Chicago group acknowledges funding for this project from the David and Lucile Packard Foundation through a fellowship to J.L.B.


\bibliography{report}   
\bibliographystyle{spiebib}   

\end{document}